\documentclass[12pt]{article}
\newcommand{\be}{\begin{equation}}
\newcommand{\ee}{\end{equation}}
\newcommand{\bea}{\begin{eqnarray}}
\newcommand{\eea}{\end{eqnarray}}
\newcommand{\nn}{\nonumber}
\newcommand{\lb}{\left[}
\newcommand{\rb}{\right]}
\newcommand{\ac}{\mathcal{A}}
\newcommand{\bac}{\bar\mathcal{A}}
\newcommand{\qc}{\mathcal{Q}}
\newcommand{\bqc}{\bar\mathcal{Q}}
\newcommand{\dc}{\mathcal{D}}
\newcommand{\bdc}{\bar\mathcal{D}}
\newcommand{\fc}{\mathcal{F}}
\newcommand{\of}{{1\over 4}}
\newcommand{\hf}{{1\over 2}}
\newcommand{\p}{\partial}

\begin{document}
\begin{titlepage}

\title{Fermions as $U(1)$ instantons}

\author{M.N. Stoilov\\
{\small\it Bulgarian Academy of Sciences,}\\
{\small\it Institute of Nuclear Research and Nuclear Energy,}\\
{\small\it Blvd. Tzarigradsko Chausse\'e 72, Sofia 1784, Bulgaria}\\
{\small e-mail: mstoilov@inrne.bas.bg}}

\maketitle

\begin{abstract}
Anomalous quantization of the electromagnetic field allows 
non-trivial (anti) self-dual configurations to exist
in four-dimensional Euclidian space-time.
These instanton-like objects are described as massless spinor particles.
\end{abstract}


key words: Electrodynamics, instanton, anomalous quantization

\end{titlepage}

Quantization in the vicinity of a fixed field configuration 
which is a solution of the equations of motion is a common technique in
Quantum Field Theory.
The classical configuration around which we quantize
 represents the vacuum of the model and the deviation of the 
field from this vacuum is the quantum field.
When the classical solution/vacuum is not unique but
depends on some parameters then a  gauge symmetry emerges in the model
\cite{t1}--\cite{t3}.
In this case the standard path integral over all fields
is replaced by functional integral (with suitable measure 
which takes into account the gauge symmetry) over the quantum field
 plus ordinary integrals over all parameters
on which the vacuum configuration depends \cite{t1}.
Here we shall consider an example where the vacuum is again 
not unique but now we do not know its explicit functional form.
Instead of this we know that it satisfies some well 
defined conditions which do not coincide with the equations of motion.  
Our example is pure Electrodynamics in  four-dimensional Euclidian 
space-time $E^4$ and the classical solution around which we decompose the 
electromagnetic potential is (anti) self-dual.
This seems rather trivial because it is known that for simple topological reasons
there are no instanton configurations in Electrodynamics. 
However, the conditions for (anti) self-duality can be rewritten  as
conditions for Fueter quaternion (anti) analyticity.
It is shown in Ref.\cite{ds2} that fields which are Fueter (anti) analytic 
can be quantized either as fermions or as bosons.
We use this result to  quantize
anomalously the (anti) self-dual electromagnetic field.
This is the key moment in the work --- we use spinors to describe instantons.
The change of the statistics `stabilizes' the instantons  allowing
non-trivial (anti) self-dual configurations to exist.
The idea that instantons change the statistics of some fields is not new 
in the non-Abelian case \cite{t4}-\cite{t6}.
Here these fields are instantons themselves.
The topological reason for  the existence of $U(1)$ instantons
is that using spinors we effectively change the
base of the $U(1)$ bundle from $E^4$ to its double covering space
which has non-trivial fundamental group.
Our conclusion is that anomaluosly quantized instanton modes of the 
electromagnetic field are described by the fermion field in massless spinor 
Electrodynamics.

The use of a non-unique classical solution as vacuum requires some precautions.
In this case the quantum evolution operator which determines the propagator of 
the quantum field (see below for the exact definition)
has zero modes.
This means that there is gauge symmetry in the resulting model,
with zero modes describing Goldstone bosons.
The existence of a zero mode is rather easy to demonstrate.
Suppose we have a model for a field $\varphi$ with Lagrangean $L$
and let $\varphi_{cl}$ is a classical solution of the equations of motion, i.e.,
\be
{\p L \over \p \varphi }\vert_{\varphi=\varphi_{cl}} = 0.\label{cls}
\ee
Let us expand the field $\varphi$ around the classical solution $\varphi_{cl}$ 
\be
\varphi=\varphi_{cl} +\eta. \label{expand}
\ee
We consider the field $\eta$ as a small  quantum variation
 of the field $\varphi$ (quantum field) and $\varphi_{cl}$ as vacuum.
Putting the expansion (\ref{expand}) for $\varphi$ into  the 
Lagrangean $L$ and keeping 
terms up to  second order  with respect to $\eta$ we get
\be
L'(\varphi_{cl},\eta)= L(\varphi_{cl}) + 
\hf\left({\p^2 L\over \p\varphi\p\varphi}\vert_{\varphi=\varphi_{cl}}\right)\eta^2.
\label{sol} 
\ee
This truncated Lagrangean is considered as a Lagrangean for the 
independent field $\eta$.
The  operator $T =
\hf(\p^2 L/ \p\varphi\p\varphi)\vert_{\varphi=\varphi_{cl}}$ 
determines the propagator of the quantum field.
This is the operator which is called `quantum evolution operator'.
Now suppose that $\varphi_{cl}$ depends on some parameter, say $\alpha$,
and let us calculate the $\alpha$-derivative of the equation of motion (\ref{cls}).
The result is
\be
0={\p\over\p\alpha}\left({\p L\over\p\varphi}\vert_{\varphi=\varphi_{cl}}\right) =
T {\p\varphi_{cl}\over\p\alpha}.
\ee
Therefore, $\p\varphi_{cl}/\p\alpha$ is a zero mode 
of the quantum evolution operator
or in other words,  there is a gauge symmetry in the Lagrangean $L'$.
This gauge symmetry emerges no matter what the initial Lagrangean $L$ is.
The symmetry has to be taken into account when we write down
the transition amplitude for the model with Lagrangean (\ref{sol}).
For this model the functional integration over the field $\varphi$ 
transforms into integral over $\alpha$ and functional integral over $\eta$
\be
\int D\varphi \;\; e^{-\int dt L} \rightarrow
\int d\alpha \int D\eta \;\; e^{-\int dt L'} \delta(\chi)\Delta\label{ch2}
\ee
where $\chi$ is some gauge fixing term and $\Delta$ is the corresponding
Faddeev--Popov determinant.
The integration over the parameter $\alpha$ effectively sums 
different classical solutions (vacuums) in the model.

It has to be stressed
 that it does not follow automatically from the above considerations
that when the Lagrangean $L$ possesses {\it ad initium} gauge symmetry 
then $L'$ has bigger gauge symmetry. 
The Lagrangean (\ref{sol}) can 
 have the same symmetry as $L$ if $\p\varphi_{cl}/\p\alpha$ coincides
with some of the zero modes due to the initial gauge freedom.
The gauge symmetry of $L'$ could be even smaller  than the symmetry of $L$
if $\varphi_{cl}$ is gauge non-invariant.
In any case one has to determine exactly the symmetry of the 
quantum evolution operator and to use adequate gauge conditions.

In the present work  we  consider a model for which
 we do not know the explicit form of $\varphi_{cl}$ and the parameters which 
it depends on, but we know that the field configuration we use as vacuum
satisfies the equation 
$
M\varphi_{cl} = 0,
$
where $M$ is some operator.
In this case eq.(\ref{ch2}) takes the form
\be
\int D\varphi \;\;  e^{-\int dt L} \rightarrow
\int D\varphi_{cl}\; \delta(M\varphi_{cl})\;
 D\eta \;\; e^{-\int dt L'} \delta(\chi)\Delta\label{ch1}. \label{ch3}
\ee
The integration over $\alpha$ in (\ref{ch2}) is replaced in (\ref{ch3}) by
$\int D\varphi_{cl} \delta(M\varphi_{cl})$ which is the way
to sum over all possible vacuums in the present case.
As a result now $\varphi_{cl}$ and $\eta$ describe two different fields.
The propagator of $\eta$ is determined as above by $T$ and
the propagator of $\varphi_{cl}$ is determined by $L(\varphi_{cl})$ and
${\rm log}({\rm det} (M))$.

The model we shall deal with hereafter is 
pure Electrodynamics  in four-di\-men\-sion\-al Euclidian space-time $E^4$.
The model is defined by the following action
\be
{\mathbf A} = - \int \of F_{\mu\nu}F_{\mu\nu}. \label{pure}
\ee
Here $F_{\mu\nu}$ is the field strength tensor for the
electromagnetic potential $A_\mu$;
 $F_{\mu\nu}=\partial_\nu A_\mu - \partial_\mu A_\nu$.
In four-dimensional space-time it is possible to use quaternions
to describe the electromagnetic field \cite{ds1}, \cite{gu}. 
The use of quaternions is especially convenient in $E^4$.

Let us remind the definition of quaternion number $\qc$
\be
\qc = q_0 + q_i e_i \label{qn}
\ee
where $q_\mu,\;\;\; \mu= 0,..,3$ are four real numbers and
$e_i,\;\;\; i=1,2,3$ are three 
non-commutative imagenary units such that:
\be
e_i e_j = - \delta_{ij} + \epsilon_{ijk} e_k. \label{qdef}
\ee
Here $\epsilon_{ijk}$ is the totally anti symmetric third rank tensor.
There is a natural operation of conjugation for quaternions.
The quaternion which is conjugated to $\qc$ we shall denote by $\bqc$
where
\be
\bqc = q_0 - q_i e_i.\label{cqn}
\ee
 
For quaternion functions there is a notion analogous to 
analyticity for complex functions --- the so called Fueter analyticity.
Let us define the first order differential operators $\bdc$ and $\dc$
\bea
\bdc&=&\partial_0 + e_i \p_i \label{fd}\\
\dc&=&\partial_0 - e_i \p_i.\label{afd}
\eea
The function $\fc$ is called Fueter analytic if it satisfies 
the following equation
\be
\bdc\fc =0.
\ee
If the function  satisfies the equation
\be
\dc\fc =0
\ee
then the function is Fueter anti analytic.

Having the electromagnetic potential  $A_\mu$ 
we can associate two quaternion functions $\ac$ and $\bac$ to it
\bea
 \ac &=&  A_0 + A_i e_i \label{qemf}\\
\bac &=& A_0 - A_i e_i. \label{cqemf}
\eea
There is also a possibility \cite{ds1} to associate a complex-valued, 
imaginary quaternion function
directly to the electromagnetic field strength tensor $F_{\mu\nu}$.
For this function it is shown in the same paper that the free Maxwell 
equations are conditions for Fueter analyticity.
Having in mind this result we want to see what the  Fueter
analyticity condition gives for the electromagnetic potential.
Using eqs.(\ref{qdef}, \ref{fd}, \ref{afd}, \ref{qemf}, \ref{cqemf}) 
we get the following result
\bea
\dc\ac &=& \partial_\mu A_\mu +
(\partial_0 A_i - \partial_i A_0 - \epsilon_{ijk}\partial_j A_k) e_i \\
\bdc\bac &=& \partial_\mu A_\mu +
(-\partial_0 A_i + \partial_i A_0 - \epsilon_{ijk}\partial_j A_k) e_i 
\eea
So, if the electromagnetic potential is such that
\be
\dc\ac =0 \label{sdu}
\ee
then we have self-dual configuration plus Lorentz gauge condition.
On the other hand if
\be
\bdc\bac =0 \label{asdu}
\ee
then the field is anti self-dual plus again the Lorentz condition.

In perturbation theory the propagator of the electromagnetic field
is Green's function of the second order Maxwell's equations.
This Green's function in momentum space behaves as $1/p^2$.
On the other hand, the instantons satisfy first order
differential equations and so their propagator in momentum space looks
like $1/p$.
Therefore perturbation theory does not take into account the 
propagation of (anti) self-dual configurations.
In order to take into account the instanton propagation we
expand  formally the electromagnetic potential $A$ around a vacuum
which is a sum of fixed self-dual potential $A^+$ and 
anti self-dual potential $A^-$, i.e. we write
\be
A_\mu = A^+_\mu + A^-_\mu + A'_\mu. \label{dec}
\ee
Then we allow $A^\pm$ to vary still remaining (anti) self-dual and
using formula (\ref{ch3}) we find the path integral measure for the instantons.
This allows us to determine their propagator.
Carrying  out this strategy we first
introduce the field strengths $F^+$, $F^-$ and $F'$ 
which correspond to the potentials $A^+$, $A^-$ and $A'$ respectively .
The sum $A^++A^-$ is a classical solution
because both $A^+$ and $A^-$ are classical solutions
and because Maxwell's equations are linear.
Using this fact and the following well-known relations
\bea
F^+_{\mu\nu}F^-_{\mu\nu}&=&0\nn\\
F^+_{\mu\nu}F^+_{\mu\nu}&=& 2 \partial_\lambda \left(
\epsilon_{\lambda\mu\nu\rho} A^+_\mu \partial_\nu A^+_\rho\right) \nn\\
F^-_{\mu\nu}F^-_{\mu\nu}&=& - 2 \partial_\lambda \left(
\epsilon_{\lambda\mu\nu\rho} A^-_\mu \partial_\nu A^-_\rho\right) \nn
\eea
it is easy to see that the electromagnetic action (\ref{pure})
in the vicinity of the field configuration $A^++A^-$  takes the form
\be
{\mathbf A}=- \int \of F'_{\mu\nu}F'_{\mu\nu}. \label{aacs}
\ee
As a consequence of  eq.(\ref{aacs}) the quantum evolution operator
in our case coincides with the Maxwell's operator. 
There are no extra zero modes and the gauge symmetry group is $U(1)$.
Certainly, in order to obtain a well-defined transition amplitude,
we have to choose a gauge fixing function.
Once again, because of eq.(\ref{aacs}) the gauge fixing function can be each 
of the standard  ones used in Electrodynamics
(now written for the field $A'$).
Therefore we can view $A'$ as the standard electromagnetic potential
 and eq.(\ref{aacs}) as the standard pure electromagnetic action.

Another consequence of eq.(\ref{aacs}) is that 
because $\mathbf A$ does not depend on $A^+$ and $A^-$ the
contribution  of (anti) instanton modes to the transition amplitude will be
a multiplicative constant which gives the number of
 different (anti) self-dual configurations.
In order to find this number we have to clarify 
what `different (anti) self-dual configurations' means,
which is the question for the gauge freedom in $A^+$ and $A^-$.
Writing eq.(\ref{dec}) we have assumed that $A'$ is a connection as well as $A$ is. 
Now, note that the difference of two connections is a tensor 
(invariant in our $U(1)$ case).
So, the proper handling of the expansion (\ref{dec}) requires to use such 
self-dual and anti self-dual potentials so that their sum $A^+ + A^-$
is gauge invariant, i.e.
\be
\delta_\epsilon (A^+_\mu + A^-_\mu)=0.\label{gv}
\ee
Here $\delta_\epsilon $ denotes the usual $U(1)$ gauge variation whose 
parameter $\epsilon$ is an arbitrary function on $E^4$.
However, even if $A^+ + A^-$ satisfies eq.(\ref{gv})
there is still room for the following gauge transformation:
\be
\delta'_\zeta (A^\pm_\mu) = \pm \partial_\mu\zeta. \label{agi}
\ee
($\zeta$ is an arbitrary function too.)
Thus the decomposition (\ref{dec}) introduces a new gauge freedom in the model
which is independent from the initial one.
But we do not want it, so we have to fix it ensuring that
\be
\delta'_\zeta (A^+_\mu - A^-_\mu)=0.\label{sgv}
\ee
So, we have to impose two conditions (\ref{gv}) and (\ref{sgv}) 
on two independent combinations of the potentials $A^+$ and $A^-$
which is equivalent to fixing the gauge separately in both the
self-dual and anti self-dual potentials.
We choose as gauge conditions 
$$\partial_\mu A^\pm_\mu =0.$$
Therefore the requirements for Fueter anti-analyticity for $\ac^+$ and 
Fueter analyticity for $\bac^-$  perfectly specify $A^+$ and $A^-$.
As a consequence the integration measure over self-dual configurations 
 which counts these configurations properly must be something like 
\be
\delta(\dc\ac^+)\;\; D A^+ \label{measure}
\ee
with an analogous expression for the anti self-dual potential.

An important note should be added at this point.
The representation of the quaternion imaginary units we are using here is
\be
e_k = -i\sigma_k, \;\;\; k=1,2,3 \label{rep}
\ee
where $\sigma_k$ are the Pauli matrices.
As a result, first,
in this representation the formulae (\ref{qemf}, \ref{cqemf})  
 can be viewed as a change of variables such that now 
 the electromagnetic potential 
has not one vector index but two spinor ones (vector transforms in bispinor).
Second, because of eqs.(\ref{rep}), any quaternion number $\qc$
 is now represented by a $2\times 2$ matrix
\be
\qc = q_0 + q_i e_i = \left(
\begin{array}{cc} 
a & - \bar b\\b & \bar a
\end{array}\right) \label{rpr}
\ee
where $a = q_0 - i q_3$, $b=q_2 - i q_1$ with $\bar a$, $\bar b$ 
denoting the complex conjugates of $a$ and $ b$.
Therefore, if for some non-zero two-dimensional  normalized vector 
(in fact, 2D normalized Weyl spinor) $v$, 
say $v=\left( \begin{array}{c} 1\\0\end{array}\right)$,
we know $\qc\cdot v $, then we know $\qc$ itself because
four real numbers $q_\mu $ are just enough to determine
two complex numbers $a$ and $b$ and vice-versa.
The same comments are by no means applicable to $\bqc$.
These arguments have to be taken into account when we write down 
the integration measure.
Using the representation (\ref{rep}) we get  
in eq.(\ref{measure}) a delta function of  a matrix which has to 
be understood as a product of delta functions of each
matrix entry.
This gives four complex conditions for $A^+$ which exceeds the 
correct number.
In order to get two complex conditions for $A^+$ 
 we have to pick up a constant spinor $v$ 
which to multiply $\dc\ac^+$ on the right and to use  $\dc\ac^+\cdot v$ 
as an argument of the delta function in eq.(\ref{measure}).
Unfortunately, following this  recipe we face a problem 
which is due to the gauge non-invariance of $v$: the quantity
$\dc\ac^+\cdot v$ is neither gauge invariant, nor gauge covariant.
So, the price we have to pay in order to have the correct number of constraints
is gauge non-invariance of the obtained expression.
We shall come back to this question later.
Let us define the spinors $\psi^+$ and $\psi^-$ as follows
\bea
\psi^+ &=& \ac^+\cdot v\nn\\
\psi^- &=& \bac^-\cdot u .\label{ano}
\eea
Then the transition amplitude for pure Electrodynamics
 plus instanton contributions and  without gauge freedom of the quantum field
 to be fixed takes the form
\be
S=\int D \psi^+ D \psi^- D A' \exp\left\{- \of\int F'_{\mu\nu}F'_{\mu\nu}\right\}
\delta(\dc\psi^+)\delta(\bdc\psi^-).\label{ta1}
\ee
Note that using spinors $\psi^+$ and $\psi^-$ to describe instantons 
and anti-instantons we effectively change the trivial $U(1)$ bundle over $E^4$ 
to a one whose base $B$ is a double cover of $E^4$.
The fundamental group of $B$ is non-zero: $\pi(B)=\mathbf Z$
which is the reason for nontrivial (anti) self dual configurations to exist.
These configurations are, however, spinors and we shall quantize them as 
Fermi particles, i.e., we suppose that $v$ and $u$ are Grassmann odd quantities.
This approach could be viewed as an inverse to the procedure used 
in Ref.\cite{ds2} for anomalous quantization of  the massless spinor field.

The two delta functions in eq.(\ref{ta1}) can be represented
as one.
We introduce four-dimensional Dirac spinor $\psi$ 
as  direct sum of Weil spinors $\psi^+$ and $\psi^-$ 
\be
\psi=\left(\begin{array}{c} 
\psi^-\\\psi^+\end{array}\right)\label{dspin}
\ee
and make use of the diagonal $\gamma^5$ representation of the 4D
 Euclidian gamma matrices
\be
\gamma_0=\left(\begin{array}{cc} 
0 & 1\\1 & 0\end{array}\right),\;\;\;
\gamma_i=\left(\begin{array}{cc} 
0 & i\sigma_i\\-i\sigma_i & 0\end{array}\right),\;\;\;
\gamma_5=\left(\begin{array}{cc} 
-1 & 0\\0 & 1\end{array}\right).
\ee
In this representation the 4D Euclidian Dirac operator has the form
\be
\partial_\mu \gamma_\mu = 
\left(\begin{array}{cc} 
0 & \partial_0 + i \partial_k \sigma_k\\
\partial_0 - i \partial_k \sigma_k& 0\end{array}\right) = 
\left(\begin{array}{cc} 
0 & \dc\\ \bdc & 0\end{array}\right). \label{dslash}
\ee
Note that the operators $\bdc$ and $\dc$ which appear in eq.(\ref{dslash})
are those defined by eqs.(\ref{fd}, \ref{afd}) taken  in the
representation (\ref{rep}).
Using eqs.(\ref{dspin}, \ref{dslash}) we get:
\be
\delta(\dc\psi^+)\delta(\bdc\psi^-)=\delta(\partial_\mu \gamma_\mu \psi).
\label{fdf}
\ee
With the help of  Lagrange multipliers the delta function (\ref{fdf})
can be represented as part of the action.
Because of the spinor character of the delta function argument and because of
hermiticity of the action, the Lagrange multipliers
have to form a spinor, conjugated to the spinor $\psi$ \cite{ds2}.
Thus we get the following expression for the transition amplitude:
\be
S=\int D\bar\psi D\psi D A' 
\exp\left\{- \int i \bar\psi \partial_\mu \gamma_\mu \psi +
\of F'_{\mu\nu}F'_{\mu\nu} \right\}.\label{ta2}
\ee
It has already been mentioned above that the introduction of the spinors
$v$ and $u$ spoils  gauge covariance
of the delta function argument.
As a result the action in eq.(\ref{ta2}) is  gauge non-invariant.
However, we have started with gauge invariant action, and we have to restore
this invariance in eq.(\ref{ta2}).
We do this in the standard way, namely prolonging the derivatives with $A'$
thus obtaining
\be
S=\int D \bar\psi D \psi D A' 
\exp\left\{- \int \bar\psi( i \partial_\mu + A'_\mu )\gamma_\mu \psi +
 \of F'_{\mu\nu}F'_{\mu\nu} \right\}.\label{ta3}
\ee
Eq.(\ref{ta3}) coincides with the expression for the transition amplitude 
of  massless spinor Electrodynamics.
In our case the spinor field represents the anomalously quantized instanton modes.
Once again, we stress that gauge is not fixed in eq.(\ref{ta1}) and
so it is not fixed in (\ref{ta3}) as well.
There is nothing special in this gauge fixing --- it is a standard
gauge fixing for the electrodynamics potential $A'$ 
and we shall not specify it explicitly.

It is possible to get eq.(\ref{ta3}) directly without
passing through eq.(\ref{ta2}).
For this purpose we have to use in eq.(\ref{ano}) a gauge invariant spinor $w$
instead of the constant but gauge non-invariant spinor $v$, where
$w$ is given by the following expression:
\be
w = e^{-i \phi\lb\Gamma\rb} v.\label{gii}
\ee
Here $\phi\lb\Gamma\rb$ is a phase which compensates the gauge
transformation of $v$.
The phase $\phi\lb\Gamma\rb$ is non-integrable \cite{d}, i.e. 
it is not a function with definete value in each space-time point $x$, 
but instead it is a multi-valued functional depending on the path $\Gamma$ along which
we reach the point $x$. 
However, the phase $\phi\lb\Gamma\rb$ possesses well defined derivatives
\be
\partial_\mu \phi\lb\Gamma\rb = A'_\mu .\nn
\ee
Formally, $\phi\lb\Gamma\rb$ can be written as an integral over 
path $\Gamma$ 
\be
\phi = \int^x_\Gamma d x_\mu A'_\mu(x).
\ee
When we use $w$ to reduce the number of 
constraints on $A^+$ we get
\bea
\dc\ac^+\cdot w & = &
\dc (e^{-i \phi\lb\Gamma\rb} \ac^+\cdot v) \nn\\
&=& e^{-i \phi\lb\Gamma\rb} (\dc - i \ac' )\psi^+ \label{gis}
\eea 
and this gauge invariant expression we have to put into  the delta function
which defines the measure over self-dual field configurations.
Representing the delta function as an exponent, the phase $ e^{-i \phi\lb\Gamma\rb}$ 
in eq.(\ref{gis}) will be compensated by the (conjugated to it) 
phase of the Lagrange multipliers.
Thus the phase $\phi\lb\Gamma\rb$ does not appear at all in the Lagrangean.
Appling the same procedure to the measure which counts anti self-dual
fields, we get for the transition amplitude directly
the gauge invariant expression (\ref{ta3}). 

Let us summarize our results:
we have started with pure Electrodynamics with  vacuum 
which is (anti) self-dual, then we quantize anomalously the vacuum modes and  
what we get finally is massless spinor Electrodynamics  where fermions
describe instanton modes.

\section*{Acknowledgement}
It is a pleasure to thank E. Nissimov and S. Pacheva
for useful discussions and to O. Stoytchev for reading the manuscript.

\section*{Dedication}
The work is dedicated to the memory of Prof. Dimitar Stoyanov

\end {document}